\documentstyle[aps,preprint,epsf]{revtex}
\newcommand{\beq}{\begin{equation}}
\newcommand{\eeq}{\end{equation}}
\begin{document}
\draft
\tightenlines

\title{ Magnetoresistance of Highly Correlated Electron Liquid}
\author{ V.R. Shaginyan$^{a,b}$\footnote{E--mail:
vrshag@thd.pnpi.spb.ru}}
\address{ $^a$Petersburg Nuclear Physics
Institute, Russian Academy of Sciences, Gatchina,
188300, Russia;\\
$^b\,$CTSPS, Clark Atlanta University,
Atlanta, Georgia 30314, USA}
\maketitle

\begin{abstract}

The behavior in magnetic fields of a highly correlated electron
liquid approaching the fermion condensation quantum
phase transition from the disordered phase
is considered. We show that at sufficiently high temperatures
$T\geq T^*(x)$ the effective mass starts to depend on $T$,
$M^*\propto T^{-1/2}$. This $T^{-1/2}$ dependence of the effective mass
at elevated  temperatures leads to the non-Fermi liquid behavior of
the resistivity, $\rho(T)\propto T$ and at higher temperatures
$\rho(T)\propto T^{3/2}$. The
application of a magnetic field $B$
restores the common $T^2$ behavior of
the resistivity.
The effective mass depends on the magnetic field,
$M^*(B)\propto B^{-2/3}$,  being approximately
independent of the temperature at $T\leq T^*(B)\propto B^{4/3}$.
At $T\geq T^*(B)$, the $T^{-1/2}$ dependence of the effective mass is
re-established. We demonstrate that this $B-T$ phase diagram has a
strong impact on the magnetoresistance (MR) of the highly correlated
electron liquid. The MR as a function of the temperature exhibits a
transition from the negative values of MR at $T\to 0$ to the positive
values at $T\propto B^{4/3}$. Thus, at $T\geq T^*(B)$,
MR as a function of the temperature possesses a node at
$T\propto B^{4/3}$.

\end{abstract}\bigskip


An explanation of the rich and striking behavior of the strongly
correlated electron liquid in heavy fermion metals and
high-temperature superconductors are, as years before, among the main
problems  of the condensed matter physics.  There is a fundamental
question about whether or not these properties can be understood
within the framework of the Landau Fermi liquid theory \cite{lan}.
The basis of the Landau theory is the assumption  that the excitation
spectrum of the Fermi liquid looks like the spectrum of an ideal
Fermi gas. This excitation spectrum is described in terms of
quasiparticles with an effective mass $M^*$, charge $e$ and spin
$1/2$. The single-particle excitations, or quasiparticles, define the
major part of the low-temperature properties of Fermi liquids.  The
stability of the ground state of a Landau liquid is determined by the
Pomeranchuk stability conditions.  The stability is violated when at
least one of the Landau effective interaction parameters becomes
negative and reaches a critical value.  The new phase at which the
stability conditions are restored is again described within the
framework of the same theory.
The Pomeranchuk conditions do not cover all the possible
instabilities.  The missed instability corresponds to the situation
when, at the temperature $T=0$, the effective mass, the most
important characteristic of Landau quasiparticles, can become
infinitely large. Such a situation, leading to profound consequences,
can take place when the corresponding Landau amplitude being
repulsive reaches some critical value.  This leads to a completely
new class of a strongly correlated Fermi liquids with the fermion
condensate (FC) \cite{ks,vol}, which is separated from that of a
normal Fermi liquid by the fermion condensation quantum phase
transition (FCQPT) \cite{ms,ars}.

As any phase transition, the quantum phase transition is driven by
a control parameter and is related
to the order parameter, which describes a broken symmetry. In our
case, the control parameter is the density $x$ of
a system and the order parameter is $\kappa({\bf p})$.
The existence of the FC state can be revealed
experimentally.
Since the order parameter $\kappa({\bf p})$ is suppressed
by a magnetic field $B$,  a weak
magnetic field $B$ will destroy the state with FC, converting the
strongly correlated Fermi liquid into the normal Landau Fermi
liquid \cite{pogsh}. In this case the magnetic field plays a role of
the control parameter. The transition from the strongly correlated
liquid into the normal Landau liquid was observed
in several experiments \cite{mac,gen,cyr,gqz}.
As soon as FCQPT occurs at  the critical point $x=x_{FC}$,
the system becomes divided into two quasiparticle
subsystems: the first subsystem is
characterized  by the quasiparticles with the effective mass
$M^*_{FC}$, while the second one is occupied by
quasiparticles with mass $M^*_L$.
The quasiparticle dispersion law in systems with FC can be
represented by two straight lines, characterized by effective masses
$M^*_{FC}$ and $M^*_L$, and intersecting near the
binding energy $E_0$.

Properties of these new quasiparticles with $M^*_{FC}$ are closely
related to the state of the system which is
characterized by the temperature
$T$, pressure or by the presence of the superconductivity.
We may say that the quasiparticle system in the
range occupied by FC becomes very ``soft''
and is to be considered as a strongly correlated liquid.
Nonetheless, the basis of the Landau Fermi liquid
theory survives FCQPT: the low energy
excitations of a strongly correlated liquid with FC are
quasiparticles.  The only difference between the Landau Fermi-liquid
and Fermi-liquid after FCQPT is that we have to expand the number of
relevant low energy degrees of freedom by introducing a new type of
quasiparticles with the effective mass $M^*_{FC}$ and the energy
scale $E_0$ \cite{ms,ams}.

When a Fermi system approaches FCQPT from the disordered phase
it remains the Landau Fermi liquid with the effective mass $M^*$
strongly depending on the density $x_{FC}-x$, temperature and
a magnetic field $B$ provided that $|x_{FC}-x|/x_{FC}\ll 1$
and $T\geq T^*(x)$ \cite{shag}.
This state of the system, with $M^*$ strongly depending on
$T$, $x$ and $B$, resembles the strongly correlated liquid.
In contrast with a strongly correlated liquid, there is no
the energy scale $E_0$ and the system under consideration is
the Landau Fermi liquid at sufficiently low temperatures with
the effective mass $M^*\simeq constant$. Therefore this liquid
can be called a highly correlated liquid.
Obviously, a highly correlated liquid has to have uncommon
properties.

In this Letter, we study the behavior of a highly correlated electron
liquid in a magnetic field. We show that at $T\geq T^*(x)$
the effective mass starts to depend on the temperature,
$M^*\propto T^{-1/2}$. This $T^{-1/2}$ dependence of the effective mass
at elevated  temperatures leads to the non-Fermi liquid behavior of
the resistivity,  $\rho(T)\sim \rho_0+aT+bT^{3/2}$. The
application of magnetic field $B$ restores the common $T^2$ behavior of
the resistivity, $\rho\simeq \rho_0+AT^2$ with  $A\propto (M^*)^2$.
Both the effective mass and coefficient $A$ depend on the magnetic
field, $M^*(B)\propto B^{-2/3}$ and $A\propto B^{-4/3}$ being
approximately independent of the temperature at $T\leq T^*(B)\propto
B^{4/3}$.  At $T\geq T^*(B)$, the $T^{-1/2}$ dependence of the
effective mass is re-established. We demonstrate that this $B-T$
phase diagram has a strong impact on the magnetoresistance (MR) of
the highly correlated electron liquid. The MR as a function of the
temperature exhibits a transition from the negative values of MR at
$T\to 0$ to the positive values at $T\propto B^{4/3}$. Thus, at
$T\geq T^*(B)$, MR as the function of the temperature possesses a
node at $T\propto B^{4/3}$. Such a behavior is of a general form and
takes place in both three dimensional (3D) highly correlated systems
and two dimensional (2D) ones.

At $|x-x_{FC}|/x_{FC}\ll 1$ and $T\to 0$, the effective mass $M^*$ of
a highly correlated electron liquid is given by the
equation \cite{shag}
\beq
M^*\sim M\frac{x_{FC}}{x-x_{FC}}.\eeq
It follows from Eq. (1) that effective mass
is finite provided that $|x-x_{FC}|\equiv \Delta x>0$. Therefore,
the system represents the Landau Fermi liquid. On the other hand,
$M^*$ diverges as the
density $x$ tends to the critical point of FCQPT. As a result,
the effective mass strongly depends on such a quantities as the
temperature, pressure, magnetic field provided that they exceed
their critical values.
For example, when
$T$ exceeds some temperature $T^*(x)$, Eq. (1) is no longer valid,
and $M^*$ depends on the temperature as well.
To evaluate this dependence, we calculate the deviation
$\Delta x(T)$ generated by $T$.
The temperature smoothing out the Fermi function $\theta(p_F-p)$
at $p_F$ induces the variation $p_F \Delta p/M^*(x)\sim T$, and
$\Delta x(T)/x_{FC}\sim M^*(x)T/p_F^2$, with $p_F$ is the
Fermi momentum and $M$ is the bare electron mass.
The deviation $\Delta x$ can be
expressed in  terms of $M^*(x)$ using Eq. (1),
$\Delta x/x_{FC}\sim M/M^*(x)$.
Comparing these deviations,
we find that at $T\geq T^*(x)$  the effective mass depends
noticeably on the temperature, and
the equation for $T^*(x)$ becomes
\beq T^*(x)\sim p_F^2\frac{M}{(M^*(x))^2}\sim \varepsilon_F(x)
\left(\frac{M}{M^*(x)}\right)^2.\eeq
Here $\varepsilon_F(x)$ is the Fermi energy of noninteracting
electrons with mass $M$. It follows from Eq. (2) that $M^*$ is
always finite at temperatures $T>0$. At $T\geq T^*(x)$, the main
contribution to $\Delta x$ comes from the temperature, therefore \beq
M^*\sim M\frac{x_{FC}}{\Delta x(T)}\sim
M\frac{\varepsilon_F}{M^*T}.\eeq As a result, we obtain \beq
M^*(T)\sim M\left(\frac{\varepsilon_F}{T}\right)^{1/2}.\eeq
Equation (4) allows us to evaluate the resistivity as a function of
$T$.  There are two terms contributing to the resistivity. Taking
into account that $A\sim (M^*)^2$ and Eq. (4), we obtain the first
term $\rho_1(T)\sim T$. The second term $\rho_2(T)$ is related to the
quasiparticle width $\gamma$. When $M/M^*\ll 1$, the width
$\gamma\propto (M^*)^3 T^2/\epsilon(M^*)\propto T^{3/2}$,
with $\epsilon(M^*)\propto (M^*)^2$ is the dielectric constant
\cite{ars,ksch}. Combining both of the contributions, we find that
the resistivity is given by
\beq \rho(T)-\rho_0\sim aT+bT^{3/2}.\eeq
Thus, it turns out that at low temperatures, $T<T^*(x)$, the
resistivity $\rho(T)-\rho_0\sim AT^2$. At higher temperatures, the
effective mass depends on the temperature and
the main contribution comes from the first term on the right hand
side of Eq. (5). At the same time, $\rho(T)-\rho_0$ follows the
$T^{3/2}$ dependence at elevated temperatures.

In the same way as Eq. (4) was derived, we can obtain the equation
determining $M^*(B)$ \cite{shag}
\beq M^*(B)\sim M\left(\frac{\varepsilon_F}
{B\mu}\right)^{2/3}, \eeq
where $\mu$ is the electron magnetic moment.
We note that $M^*$ is determined by Eq. (6) as long as
$M^*(B)\leq M^*(x)$, otherwise we have to use Eq. (1).
It follows from Eq. (6) that the application of a magnetic field
reduces the effective mass.  Note, that if there exists an itinerant
magnetic order in the system which is suppressed by magnetic field
$B=B_{c0}$, Eq. (6) has to be replaced by the equation \cite{pogsh},
\beq M^*(B)\propto \left(\frac{1}
{B-B_{c0}}\right)^{2/3}.\eeq
The coefficient of $T^2$ in the expression for the resistivity
$A(B)\propto (M^*(B))^2$ diverges as
\beq A(B)\propto \left(\frac{1}
{B-B_{c0}}\right)^{4/3}.\eeq
At elevated temperature, there is a temperature $T^*(B)$ at which
$M^*(B)\simeq M^*(T)$. Comparing Eq. (4) and Eq. (7), we see that
$T^*(B)$ is given by
\beq T^*(B)\propto (B-B_{c0})^{4/3}. \eeq
At $T\geq T^*(x)$, Eq. (9) determines the line in the $B-T$ phase
diagram which separates the region of
the $B$ dependent effective mass from the
region of the $T$ dependent effective mass.
At the temperature $T^*(B)$, there occurs a
crossover from the $T^2$ dependence
of the resistivity to the $T$ dependence: at $T<T^*(B)$,
the effective mass is given by Eq. (7), and at $T>T^*(B)$
$M^*$ is given by Eq. (4).

Using the $B-T$ phase diagram just described above,
we consider the behavior of MR
\beq \rho_{mr}(B,T)=\frac{\rho(B,T)-\rho(0,T)}{\rho(0,T)},\eeq
as a function of magnetic
field $B$ and $T$. Here $\rho(B,T)$ is the
resistivity measured at the magnetic field
$B$ and temperature $T$. We assume that the contribution
$\Delta\rho_{mr}(B)$
coming from the magnetic field $B$ can be treated within
the low field approximation and given by the
well-known Kohler's rule,
\beq\Delta\rho_{mr}(B)\sim B^2\rho(0,\Theta_D)/\rho(0,T),\eeq
with $\Theta_D$ is the Debye temperature. Note, that the low
field approximation implies that
$\Delta\rho_{mr}(B)\ll \rho(0,T)\equiv\rho(T)$.
Substituting Eq. (11)
into Eq. (10), we find that
\beq \rho_{mr}(B,T)\sim
\frac{c(M^*(B,T))^2T^2+\Delta\rho_{mr}(B)-c(M^*(0,T))^2T^2}
{\rho(0,T)}.\eeq
Here $M^*(B,T)$ denotes the effective mass $M^*$ which now depends
on both magnetic field and the temperature, and $c$ is a constant.

Consider MR given by Eq. (12) as a function
of $B$ at some temperature $T=T_0$. At low temperatures
$T_0\leq T^*(x)$, the system behaves as common Landau Fermi liquid,
and MR is an increasing function of $B$. When the
temperature $T_0$ is sufficiently high, $T^*(B)<T_0$, and
the magnetic field is  small, $M^*(B,T)$ is given by Eq. (4).
Therefore, the difference $\Delta M^*=|M^*(B,T)-M^*(0,T)|$ is small
and the main contribution is given by $\Delta\rho_{mr}(B)$.  As a
result, MR is an increasing function of $B$. At elevated $B$, the
difference $\Delta M^*$ becomes a decreasing function of $B$, and MR
as the function of $B$ reaches its maximum value at $T^*(B)\sim T_0$.
In accordance with Eq. (9), $T^*(B)$ determines the crossover from
$T^2$ dependence of the resistivity to the $T$ dependence.
Differentiating the function $\rho_{mr}(B,T)$ given by Eq.
(12) with respect
to $B$, one can verify that the derivative
is negative at sufficiently large values
of the magnetic field when $T^*(B)\simeq T_0$.
Thus, we are led to the conclusion that the crossover manifests
itself as the maximum of MR as the function of $B$.

We now consider MR as a function of $T$ at some $B_0$.
At low temperatures $T\leq T^*(x)$, we have a normal Landau liquid.
At higher temperatures $T^*(x)<T\ll T^*(B_0)$, it follows
from Eqs. (4) and (7) that $M^*(B_0)/M^*(T)\ll 1$, and
MR is determined by the resistivity $\rho(0,T)$.
Note, that $B_0$ has to be comparatively
high to ensure the inequality,
$T^*(x)\leq T\ll T^*(B_0)$.
As a result,
MR tends to $-1$, $\rho_{mr}(B_0,T\to0) \sim -1$.
Differentiating the function
$\rho_{mr}(B_0,T)$
with respect to $B_0$ we can check that its slope becomes steeper
as $B_0$ is decreased, being proportional
$\propto (B_0-B_{c0})^{-7/3}$.
At $T=T_1\sim T^*(B_0)$, MR possesses a node because at this
point the effective mass $M^*(B_0)\simeq M^*(T)$,
and $\rho(B_0,T)\simeq \rho(0,T)$.
Again, we can conclude that the crossover from the $T^2$
resistivity to the $T$ resistivity, which occurs at $T\sim T^*(B_0)$,
manifests itself in
the transition from negative MR to positive MR.
At $T>T^*(B_0)$,
the main contribution
in MR comes from $\Delta\rho_{mr}(B_0)$, and
MR reaches its maximum value.
Upon using Eq. (11) and taking into account that at this point
$T$ has to be determined by Eq. (9), $T\propto (B_0-B_{c0})^{4/3}$,
we obtain that the maximum value
$\rho^m_{mr}(B_0)$ of MR is
$\rho^m_{mr}(B_0)\propto (B-B_{c0})^{-2/3}$.
Thus, the maximum value is a decreasing function of $B_0$.
At $T^*(B_0)\ll T$, MR is a decreasing function of the temperature,
and at elevated temperatures MR eventually vanishes since
$\Delta\rho_{mr}(B_0)/\rho(T)\ll 1$.

The recent paper \cite{pag} reports on measurements of
the resistivity of CeCoIn$_5$ in
a magnetic field. With increasing field, the resistivity evolves
from the $T$ temperature dependence to the $T^2$ dependence, while
the field dependence of $A(B)\sim (M^*(B))^2$ displays the critical
behavior best fitted by the function, $A(B)\propto
(B-B_{c0})^{-\alpha}$, with $\alpha\simeq 1.37$ \cite{pag}.  All the
data are in a good agreement with the $B-T$ phase diagram given by
Eq. (9). The critical behavior displaying $\alpha=4/3$ \cite{shag}
and described by Eq. (8) is also in a good agreement with the data.
Transition from negative MR to positive MR with increasing
$T$ was also observed \cite{pag}. We believe that an additional
analysis of the data \cite{pag} can reveal that the crossover
from $T^2$ dependence of the resistivity to the $T$ dependence
occurs at $T\propto (B-B_{c0})^{4/3}$. As well, this analysis
could reveal supplementary peculiarities of MR.

In conclusion,  we have described the behavior of a highly correlated
electron liquid in a magnetic field. The highly correlated liquid
exhibits the strong dependence of the effective mass $M^*$ on the
temperature and the magnetic field. This strong dependence is of a
crucial importance when describing the $B-T$ phase diagram and such
properties as MR and the critical behavior. We have also identified
the behavior of the heavy fermion metal CeCoIn$_5$ in magnetic fields
displayed in Ref. \cite{pag} as the highly correlated behavior of a
Landau Fermi liquid approaching FCQPT from the disordered phase.

I am grateful to CTSPS for the hospitality during
my stay in Atlanta. I also thank G. Japaridze for fruitful
discussions.  This work was supported in part by the Russian
Foundation for Basic Research, No 01-02-17189.

\end{document}